\documentclass[conference]{IEEEtran}

\IEEEoverridecommandlockouts

\usepackage[cmex10]{amsmath}
\usepackage{array}
\usepackage{mdwmath}
\usepackage{mdwtab}
\usepackage{eqparbox}
\usepackage{amsmath}
\usepackage{amssymb,amsbsy}
\usepackage{amsthm}
\usepackage{graphicx}
\usepackage{color}
\usepackage{algorithm}
\usepackage{algpseudocode}
\usepackage{multirow}

\newcommand{\Fig}[1]{Fig.~\ref{#1}}
\graphicspath{{figures/}}

\begin{document}

\sloppy

\title{On Multilevel Coding Schemes Based on Non-Binary LDPC Codes \thanks{The research was carried out at the IITP RAS and supported by the Russian Science Foundation (project no. 14-50-00150).}}

\author{
\IEEEauthorblockN{Valeriya Potapova\IEEEauthorrefmark{1}\IEEEauthorrefmark{2}\IEEEauthorrefmark{3} and Alexey Frolov\IEEEauthorrefmark{1}\IEEEauthorrefmark{2}}

\IEEEauthorblockA{\small \IEEEauthorrefmark{1} Skolkovo Institute of Science and Technology\\
Moscow, Russia
}
\IEEEauthorblockA{\small \IEEEauthorrefmark{2} Institute for Information Transmission Problems\\
Russian Academy of Sciences\\Moscow, Russia
}

{valeriya.potapova@skolkovotech.ru, al.frolov@skoltech.ru}
}




\maketitle
\begin{abstract}
 We address the problem of constructing of coding schemes for the channels with high-order modulations. It is known, that non-binary LDPC codes are especially good for  such channels and significantly outperform their binary counterparts. Unfortunately, their decoding complexity is still large. In order to reduce the decoding complexity we consider multilevel coding schemes based on non-binary LDPC codes (NB-LDPC-MLC schemes) over smaller fields. The use of such schemes gives us a reasonable gain in complexity. At the same time the performance of NB-LDPC-MLC schemes is practically the same as the performance of LDPC codes over the field matching the modulation order. In particular by means of simulations we showed that the performance of NB-LDPC-MLC schemes over GF($16$) is the same as the performance of non-binary LDPC codes over GF($64$) and GF($256$) in AWGN channel with QAM~$64$ and QAM~$256$ accordingly. We also perform a comparison with binary LDPC codes.
\end{abstract}

\section{Introduction}

In this paper we address the problem of constructing of coding schemes for the channels with high-order modulations.  Using low-density parity-check (LDPC) codes \cite{G, T} with a field order equal to the size of the constellation ($M$) has a clear advantage: modulation and demodulation are very simple and there is no loss of performance due to a demodulation at the receiver. At the same time non-binary LDPC (NB LDPC) have advantages over binary LDPC codes. Davey and MacKay \cite{DM} were first who used belief propagation (BP) to decode such codes. They showed that NB LDPC codes significantly outperform their binary counterparts. Moreover, non-binary LDPC codes are especially good for the channels with burst errors and high-order modulations \cite{SC}. 

In this paper we apply multilevel coding schemes based on non-binary LDPC codes over small fields. Multilevel coding was proposed in \cite{U, IH}. In \cite{WFH} MLC based on binary codes was investigated, and it was shown that binary codes can be sufficiently effective in achieving capacity with MLC, if a well suited design of the rates of the component codes  is applied. In \cite{BSS} MLC based on binary LDPC codes was investigated. In \cite{DCG} different NB LDPC modulations for higher order QAM constellations were considered and it was shown, that the bigger the alphabet size, the closer the performance curve to the Shannon limit. Here we pursue another aim: we want to reduce the complexity as much as possible and to leave the performance curve practically the same.

Our contribution is as follows. We show that the performance of NB-LDPC-MLC schemes over GF($16$) is the same as the performance of NB LDPC codes over GF($64$) and GF($256$) in AWGN channel with QAM~$64$ and QAM~$256$ correspondingly. At the same time the use of such schemes gives us a reasonable gain in complexity. We also perform a comparison with binary LDPC codes.

The paper is organized in the following way: in Section II we briefly introduce non-binary LDPC codes, and Sum-Product decoding algorithm. In Section III we provide the proposed scheme, and in Section IV simulation results are presented. Finally in Section V we analyze complexity of the proposed scheme. 

\section{Preliminaries}

\subsection{LDPC codes over GF($q$)}
An LDPC code $\mathcal{C}$ of length $N$ over GF($q$) is a null-space of  an $M \times N$ sparse parity-check matrix $\mathbf{H} = \left[ h_{i,j} \right]$, $1 \leq i \leq M$, $1 \leq j \leq N$, over GF($q$). By $\ell_j$, $j = 1, \ldots, N$, we denote the weight of $j$-th column, by $\Delta_i$, $i = 1, \ldots, M$, we denote the weight of $i$-th row. Here and in what follows by weight we mean the Hamming weight, i.e. the number of non-zero elements in a vector. By $\overline\Delta$ and $\overline\ell$ we denote average row and column weights, by $\Delta_\text{max}$ we denote maximal row weight in the parity-check matrix.The following inequality follows for the rate of the code $\mathcal{C}$
\[
R(\mathcal{C}) \geq 1 - \frac{\overline\ell}{\overline\Delta},
\] 
the equality takes place in case of full rank of $\mathbf{H}$.

In what follows we need the following notations:
\[
\Gamma(j) = \left\{i: h_{i,j} \ne 0, 1 \leq i \leq M \right\} 
\]
and
\[
\Phi(i) = \left\{ j: h_{i,j} \ne 0, 1 \leq j \leq N \right\}. 
\]

\subsection{Tanner graph}
\begin{figure}[htbp]
\centering
\includegraphics[width=0.9\linewidth]{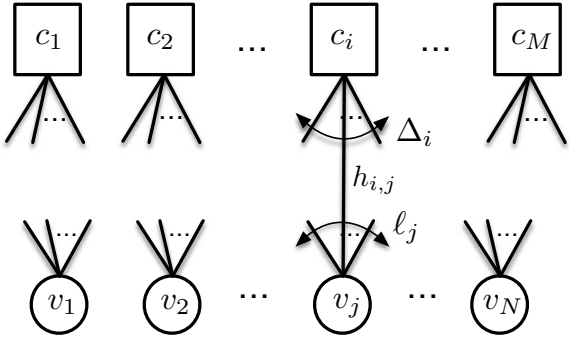}
\caption{Tanner graph}
\label{tanner}
\end{figure}

The constructed code $\mathcal{C}$ can be described with use of bipartite graph, which is called the Tanner graph \cite{T} (see \Fig{tanner}). The vertex set of the graph consists of the set of variable nodes $V = \{v_1, v_2, \ldots, v_N\}$ and the set of check nodes $C = \{c_1, c_2, \ldots, c_M\}$. The variable node $v_j$ and the check node $c_i$ are connected with an edge if and only if $h_{i,j} \ne 0$. The edge has a label $h_{i,j}$ (see \Fig{tanner}).

To check if $\mathbf{r} = (r_1, r_2, \ldots, r_N) \in \text{GF}(q)^N$ is a codeword of $\mathcal{C}$ we associate the symbols of $\mathbf{r}$ to the variable nodes ($v_j \gets r_j, j = 1,\ldots, N$). Each check node $c_i$, $1 \leq i \leq M$ imposes the following linear restriction
\[
c_i: \sum\limits_{t \in \Phi(i)} h_{i,t} r_t = 0
\]
and we can say, that linear $[\Delta_i, \Delta_i-1]$ single parity-check (SPC) codes over GF($q$) are associated to the check nodes. In what follows we refer the codes as component codes. The word $\mathbf{r}$ is a codeword of $\mathcal{C}$ if all the component codes are satisfied (the symbols which come to the codes via the edges of the Tanner graph form codewords of the component codes). 

\subsection{Sum-Product decoding algorithm}

Davey and MacKay \cite{DM} were first who used Sum-Product algorithm (SPA) to decode NB LDPC codes. The algorithm introduced in \cite{DM} (in what follows we refer the algorithm as Q-ary SPA or QSPA) is a generalization of binary Sum-Product algorithm \cite{KFL}. QSPA was defined in probability domain, log-domain decoding was suggested in \cite{WSM}. The computation complexity can be further reduced. The idea of using a multi-dimensional Fast Fourier transform (FFT) in the QSPA was proposed in \cite{BD}. The exact description of this algorithm in the log-domain is given in \cite{SC}. In \cite{DF}, the the algorithm was described with use of a tensor representation of messages. With this representation, the generalization of QSPA over GF($2$) to any field of order GF($2^m$) becomes very natural.

Analogous to the binary case, QSPA is a message passing algorithm. The main difference is that in non-binary case the messages (which go through the edges of the Tanner graph) are no more real values but the vectors of real values. In what follows we consider only QSPA with use of FFT, we refer the algorithm as FFT-QSPA. Let us first briefly explain the algorithm, for simplicity we consider probability domain only.

\textbf{Input and messages.}
Suppose we send a code word $\mathbf{c}=({{c}_{1}},{{c}_{2}},\ldots ,{{c}_{N}})\in GF{{(q)}^{N}}$ and receive a word $\mathbf{y}=({{y}_{1}},{{y}_{2}},\ldots ,{{y}_{N}}) \in \mathbb{C}^N$. Let us introduce a notation of a distribution vector
\[
\mathbf{D} = (\mathbf{d} _1, \mathbf{d} _2, \ldots, \mathbf{d}_N),
\]
where
\begin{gather*}
\begin{split}
\mathbf{d} _i = &( \Pr(c_i = \alpha_1 | y_i), \\
                            & \Pr(c_i = \alpha_2 | y_i), \\
                            &  \ldots, \\
                            & \Pr(c_i = \alpha_q | y_i)), \quad 1 \leq i \leq N.
\end{split}
\end{gather*}

The input of the algorithm is the a priori distribution vector
\[
{{\mathbf{D}}^{(0)}}=({\mathbf{d}}^{(0)}_{1}, {\mathbf{d}}^{(0)}_{2},\ldots ,{\mathbf{d}}^{(0)}_{N}),
\] 
calculated from the received vector $\mathbf{y}=({{y}_{1}},{{y}_{2}},\ldots ,{{y}_{N}})$. At each iteration the FFT-QSPA updates the vector (we denote by ${{\mathbf{D}}^{(t)}}$ the distribution vector after iteration $t$).

Analogous to the binary case, each iteration can be divided into two stages: check nodes update and variable node update. Let us consider these steps in more detail. As usual by $Q_{j \to i}$ we denote the message from $j$-th variable node to $i$-th check node and by $R_{i \to j}$ we denote the message from $i$-th check node to $j$-th variable node.

\textbf{Check node update}
Let us consider the processing of messages in the $i$-th check node. This node has degree ${{\Delta}_{i}}$. For each variable node connected to the check node we calculate a new distribution, using the current ones. 
\[
R_{i \to j} = h_{i,j}^{-1}\mathop{\otimes}\limits_{k \in \Phi(i), k \ne j} h_{i, k} Q_{k \to i}, \: \: j \in \Phi(i),
\]
where $\mathop{\otimes}$ denotes a convolution of distributions.

Recall, that the direct implementation of a convolution of two distributions requires  ${{q}^{2}}$ operations. At the same time the complexity of a convolution can be significantly reduced ($q\log q$ operations) by means of multi-dimensional FFT.

\textbf{Variable node update.}
During the variable node update we need to calculate the element-wise product of distributions, coming to the variable node:
\[
Q_{j \to i} = \mathbf{d}_j^{(0)} \odot \left(  \mathop{\odot}\limits_{k \in \Gamma(j), k \ne i} R_{k \to j} \right), i \in \Gamma(j),
\]
where $\mathop{\odot}$ denotes the element-wise product of distributions.

\section{Multilevel Coding Schemes Based on Non-Binary LDPC Codes}

For high order modulations (starting with QAM 64) we suggest to use multilevel coding (MLC) schemes \cite{U, IH} based on NB LDPC codes. In this section, we will explain the construction and decoding algorithms.

\begin{figure}[h]
\centering
\includegraphics[width=0.9\linewidth]{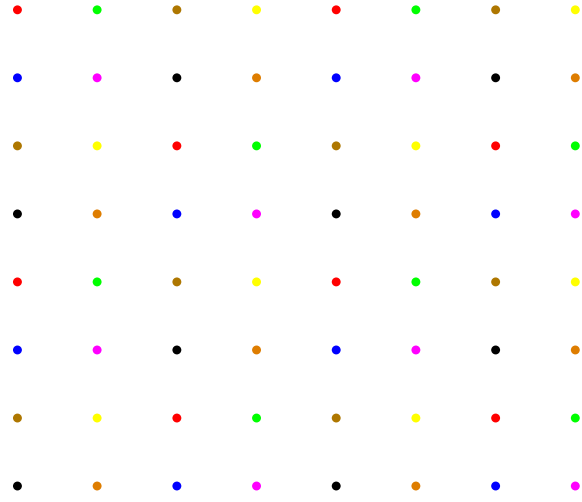}
\caption{Cosets, QAM 64}
\label{fig:cosets}
\end{figure}

To explain the construction we consider the QAM~$64$ modulation. With use of this modulation, we send $6$ bits per channel use, in other words, vector of length 6. Let us split the elements of the vector into parts (e.g. $3$ and $3$ bits) and use NB LDPC codes $\mathcal{C}_0$ and $\mathcal{C}_1$ of different rates $R_0$ and $R_1$ to protect them. In more details, suppose we want to transmit the point from Fig.2. The first $3$ bits correspond to the coset to which transmitted point belongs, and the last $3$ bits correspond to the color of the point. Let us protect the first $3$ bits with a lower rate code and the second $3$ bits -- with higher rate code (i.e. $R_0 < R_1$). It is important now to split QAM~$64$ constellation into sub-constellations (cosets) with unequal error rates. For example in \Fig{fig:cosets} we show how to do it for $(3, 3)$ case. Each color corresponds to a sub-constellation.   

\begin{figure}[htbp]
\centering
\includegraphics[width=0.9\linewidth]{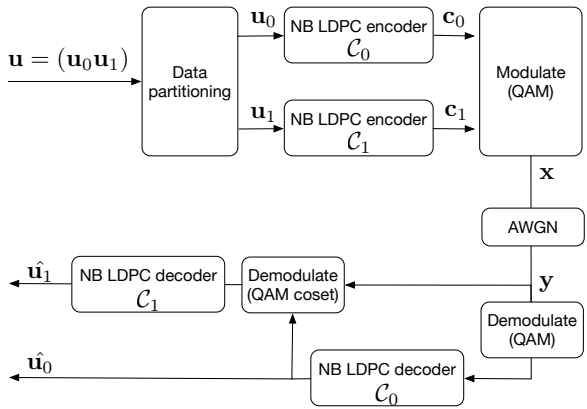}
\caption{Multilevel coding based on NB LDPC codes}
\label{fig:mlc}
\end{figure}

In \Fig{fig:mlc} the multilevel coding (MLC) scheme is shown. $u_0$ is the first $3$ bits, and $u_1$ is the last $3$ bits. We code $u_0$ and $u_1$ with NB LDPC codes $\mathcal{C}_0$ and $\mathcal{C}_1$ correspondingly, modulate with QAM 64, and transmit through the channel. The receiver then demodulates with QAM 64 vector, and  decode the first part of the vector. After that the receiver demodulates cosets with QAM 64, and finally decode the output with $\mathcal{C}_1$. As a result we have joint coding-modulation technique.

It is worth to note, that we have here two-level scheme. For $n$-level scheme the construction is the same.

\section{Simulation results}

In this section we give simulation results. We consider AWGN channel with QAM~$64$ and QAM~$256$ modulations. The code parameters are as follows: the code length is $12000$ bits, $R = 0.8$, we used FFT-QSPA with $30$ iterations. We constructed all the codes with use of Progressive Edge Grows (PEG) algorithm \cite{XEA}. For binary LDPC codes we used optimal column weigth distribution obtained by means of Density Evolution method. In case of NB LDPC codes all the columns have weight $2$. 

\subsection{QAM~$64$}

\begin{figure}[h]
\centering
\includegraphics[width=2.5in,height=2in]{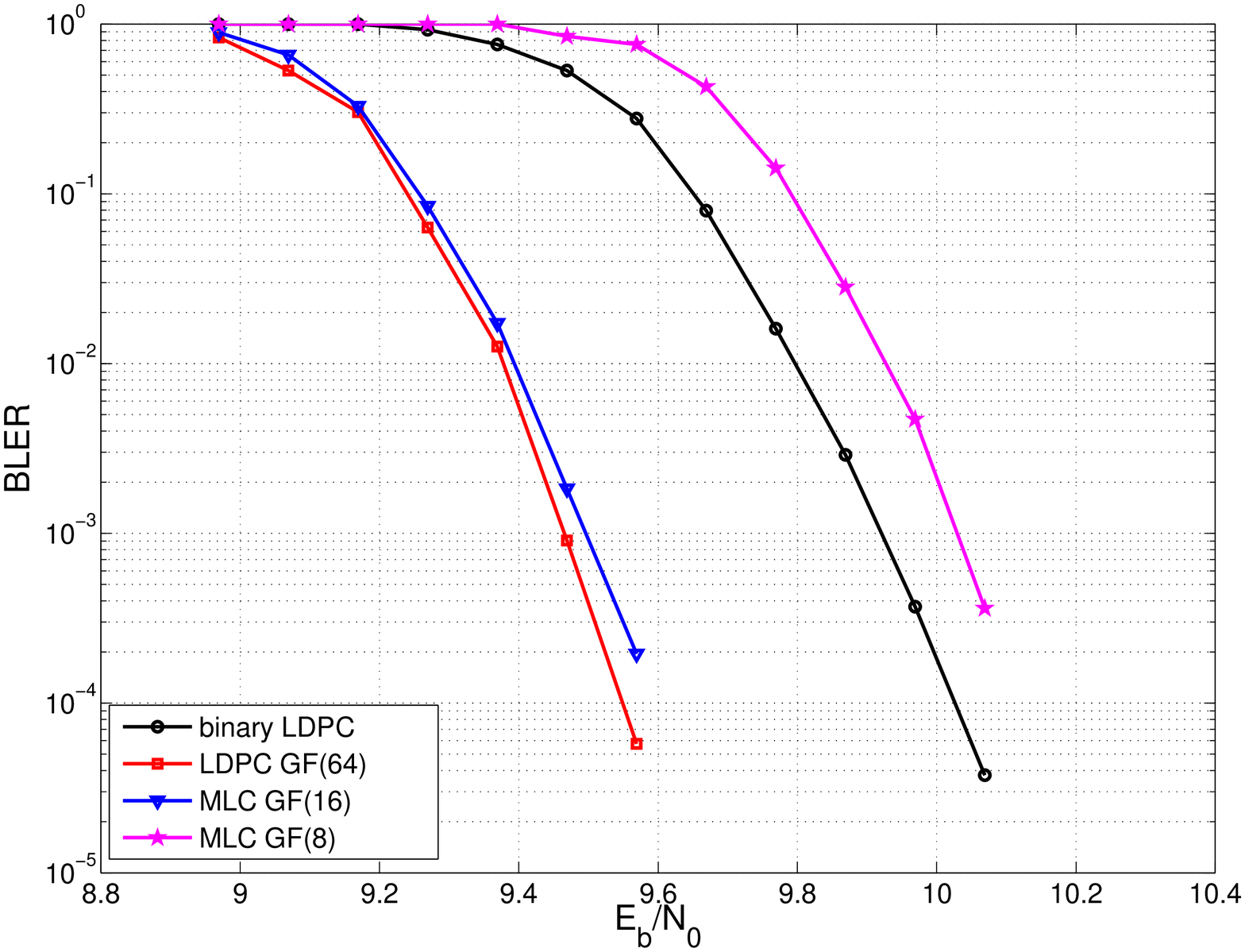}
\caption{QAM~$64$, Block error rate (BLER) vs $E_b/N_0$}\label{fig:mlcres64}
\end{figure}

Four curves are presented in \Fig{fig:mlcres64}, let us consider them in more detail:
\begin{enumerate}
\item binary LDPC codes, Gray mapping;
\item LDPC code over GF($64$), field matches the constellation order, Gray mapping;
\item MLC over GF($16$): $4$ bits are protected with $(2000, 1400)$ LDPC code over GF($16$) ($R_0 = 0.7$), $2$ bits are uncoded ($R_1 = 1$) ;
\item MLC over GF($8$): $3$ bits are protected with $(2000, 1300)$ LDPC code over GF($8$) ($R_0 = 0.65$), $3$ bits are protected with $(2000, 1900)$ LDPC code over GF($8$) ($R_1 = 0.95$);
\end{enumerate}

We see, that NB LDPC code over $GF(64)$ is approx. $0.45$ dB better, than binary LDPC code (at the level BLER = $10^{-3}$). It should be mentioned that binary LDPC code was tuned specially for good performance. Note, that MLC scheme over GF($8$) is rather bad, so GF($8$) field is not enough for QAM~$64$. At the same time the performance of MLC over $GF(16)$ is very close to the performance of pure LDPC over $GF(64)$, these two curves practically coincide.
It should be mentioned, that we have calculated the Shannon Limit for this case, and it is equal to $8.61$ dB.

\subsection{QAM~$256$}

\begin{figure}[h]
\centering
\includegraphics[width=2.5in,height=2in]{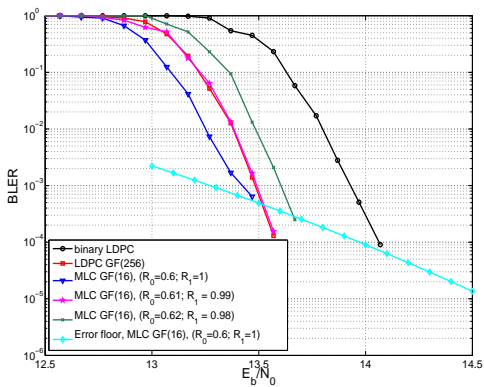}
\caption{QAM~$256$, Block error rate (BLER) vs $E_b/N_0$}\label{fig:mlcres256}
\end{figure}

Five curves are presented in \Fig{fig:mlcres256}, let us consider them in more detail:
\begin{enumerate}
\item binary LDPC codes, Gray mapping;
\item LDPC code over GF($256$), field matches the constellation order, Gray mapping;
\item MLC over GF($16$): $4$ bits are protected with $(1500, 900)$ LDPC code over GF($16$) ($R_0 = 0.6$), $4$ bits are uncoded ($R_1 = 1$) ;
\item MLC over GF($16$): $4$ bits are protected with $(1500, 915)$ LDPC code over GF($16$) ($R_0 = 0.61$), $4$ bits are protected with $(1500, 1485)$ LDPC code over GF($16$) ($R_1 = 0.99$);
\item MLC over GF($16$): $4$ bits are protected with $(1500, 930)$ LDPC code over GF($16$) ($R_0 = 0.62$), $4$ bits are protected with $(1500, 1470)$ LDPC code over GF($16$) ($R_1 = 0.98$);
\end{enumerate}

We see, that NB LDPC code over $GF(64)$ is approx. $0.4$ dB better, than binary LDPC code (at the level BLER = $10^{-3}$). In the first MLC we tried to leave $4$ bits uncoded, but this leads to high error floor (but this scheme has the best waterfall). It should be explained in more details, how we estimated the error floor for MLC $GF(16)$ ($R_0=0.6$ ,$R_1=1$). Suppose that the first bits are decoded without errors. It means, that even if there were errors in the first part of the message, the code has corrected them. So the probability that the extracted message is erroneous is the same as the probability that at least one error has occurred in the second (uncoded) part of the message.

In the rest two schemes we used the codes of rates $0.99$ and $0.98$ to protect these bits.  Again we note, that the performance curves of MLC over $GF(16)$ (scheme 4) and LDPC over $GF(256)$ coincide. For this case we also have calculated the Shannon limit, and it is equal to $12.07$ dB.

\section{Complexity analysis}

In this section we perform the complexity analysis and show that NB-LDPC-MLC schemes give a reasonable gain in complexity in comparison to  LDPC codes over the field matching the modulation order ($q=M$)\cite{BD}. As we have mentioned, we used here Q-ary Sum-Product algorithm with FFT,
\\
\\
{\bf Complexity of one iteration (NB-LDPC GF($q$)):}

\begin{table}[h]
\small
\centering
\begin{tabular}{l|l}

$+$, $GF(q)$ & --- \\
$*$, $GF(q)$ & $2 (1-R) N \overline\Delta$ \\
$+$, float & $2 (1-R) N \overline\Delta q \log_2 q$ \\
$*$, float & $N (1-R) (2 \overline\Delta -1)(q-1) + N (2 \overline\ell -1)(q-1)$ \\
memory & $(1-R) N \Delta_\text{max} (q-1)$

\end{tabular}
\end{table}

In Table~\ref{tab:compl} we give the exact number of operations and memory consumption for the coding schemes considered above. We see, that NB-LDPC-MLC is much simpler in implementation in comparison to LDPC codes with $q=M$. We recall, that the performance curves of these two schemes coincide. In other words, the proposed scheme shows the same performance and is much more simple than usual LDPC.

\begin{table}[h]
\scriptsize
\begin{tabular}{|l||c|c|c|c|c|c|c|c|}
& \multicolumn{2}{c}{QAM~$64$} & \multicolumn{2}{c}{QAM~$256$} \\
& LDPC GF($64$) & MLC GF($16$) & LDPC GF($256$) & MLC GF($16$) \\ 
\hline
$*$, $GF(q)$ & 8000 & 9000 & 6000 & 6750 \\
$+$, float & 3072000 & 576000 & 12288000 & 432000 \\
$*$, float & 856800 & 231000 & 2601000 & 171000 \\
memory & 277200 & 72000 & 841500 & 54000\\
\end{tabular}
\caption{Complexities of one iteration}
\label{tab:compl}
\end{table}

\section{Conclusion}

Here we summarize the advantages of MLC approach. We showed that the performance of NB-LDPC-MLC schemes over GF($16$) is the same as the performance of NB LDPC codes over GF($64$) and GF($256$) in AWGN channel with QAM~$64$ and QAM~$256$ accordingly. At the same time the use of such schemes gives us a reasonable gain in complexity.

\section*{Acknowledgment} 
The authors would like to thank Victor Zyablov for recommendations and advice, which were helpful in improving the paper.





\end{document}